\documentclass[amsmath,amssymb,twocolumn]{revtex4}
\usepackage[parfill]{parskip}    
\usepackage{graphicx}
\usepackage{amssymb}
\usepackage{epstopdf}
\usepackage{color}
\usepackage{graphicx}
\usepackage{pdfpages}
\usepackage{soul}
\usepackage{amsmath}

\begin{document}

\title{Four dimensional Einstein-power-Maxwell black hole solutions \\
in scale-dependent gravity}
\author{
\'Angel Rinc\'on {${}^{a}$
\footnote{angel.rincon@pucv.cl}
}   
Ernesto Contreras {${}^{b}$
\footnote{econtreras@usfq.edu.ec}
}
Pedro Bargue\~no {${}^{c}$
\footnote{pedro.bargueno@ua.es}
}
Benjamin Koch {${}^{d}$
\footnote{bkoch@fis.puc.cl}
}
Grigoris Panotopoulos  {${}^{e}$
\footnote{grigorios.panotopoulos@tecnico.ulisboa.pt}
}
}
\address{
${}^a$ Instituto de F\'isica, Pontificia Universidad Cat\'olica de Valpara\'iso, Avenida Brasil 2950, Casilla 4059, Valpara\'iso, Chile.
\\
${}^b$ Departamento de F\'isica, Colegio de ciencias e Ingenier\'ia, Universidad San Francisco de Quito, Quito, Ecuador.
\\
${}^c$ Departamento de F\'{i}sica Aplicada, Universidad de Alicante, Campus de San Vicente del Raspeig, E-03690 Alicante, Spain.
\\
${}^d$ Instituto de F{\'i}sica, Pontificia Universidad Cat{\'o}lica de Chile, Av. Vicu{\~n}a Mackenna 4860, Santiago, Chile;\\
Institut f\"ur Theoretische Physik,
 Technische Universit\"at Wien,
 Wiedner Hauptstrasse 8-10,
 A-1040 Vienna, Austria.
\\
${}^e$ Centro de Astrof{\'i}sica e Gravita{\c c}{\~a}o, Instituto Superior T{\'e}cnico-IST, Universidade de Lisboa-UL, Av. Rovisco Pais, 1049-001 Lisboa, Portugal.   
}

\begin{abstract}
In the present work, we extend and generalize our previous work regarding the scale dependence applied to black holes in the presence of non-linear electrodynamics \cite{SD2}. 
The starting point for this study is the Einstein-power-Maxwell theory with a vanishing cosmological constant in (3+1) dimensions, assuming a scale dependence of both the gravitational and the electromagnetic coupling. We further examine the corresponding thermodynamic properties and how these quantities experience deviations from their classical counterparts. 
We solve the effective Einstein's field equations using the ``null energy condition" to obtain analytical solutions. The implications of quantum corrections are also briefly discussed. Finally, we analyze our solutions and compare them to related results in the literature.  
\end{abstract}

\maketitle

\section{Introduction}

Einstein's theory of General Relativity (GR) \cite{GR}, a geometric theory of gravity compatible with Special Relativity, is not only beautiful but also very successful as well \cite{tests1,tests2}. Indeed, both the classical and solar system tests \cite{tests3}, and a few years back the direct detection of gravitational waves by the aLIGO/VIRGO observatories \cite{ligo1} have confirmed a series of remarkable predictions of GR, including the existence of gravitational waves. In fact, a series of additional gravitational wave events from black hole mergers \cite{ligo2,ligo3,ligo4,ligo5}, combined with the first image of a black hole from the Event Horizon Telescope last year \cite{L1,L2,L3}, have provided us with the strongest evidence so far that black holes (BHs) exist in nature.
\\
Despite the phenomenological success of classical GR 
there are numerous open questions concerning the quantum nature of this theory. The quest for a theory of gravity that consistently incorporates quantum mechanics is still one of the major challenges in modern theoretical physics. Most current approaches to the problem found in the literature (for a partial list, see e.g., \cite{QG1,QG2,QG3,QG4,QG5,QG6,QG7,QG8,QG9} and references therein), seem to share one particular property. The couplings that enter into the action defining ones favorite model, such as the cosmological constant, the gravitational and electromagnetic couplings etc, become scale-dependent (SD) quantities at the level of an effective averaged action after incorporating quantum effects. 
This was to be expected in some sense that since scale dependence at the level of the effective action is a generic feature of ordinary quantum field theory.
\\
On the other hand, classical electrodynamics is based on a system of linear Maxwell equations. However, as is usual in quantum physics, the effective equations become non-linear when quantum effects are taken into account. Many decades back, and in particular in the 30's, Euler and Heisenberg calculated QED corrections \cite{Euler}, while Born and Infeld managed to obtain a finite self-energy of point-like charges \cite{BI}. Those works triggered the interest in non-linear electrodynamics (NLE), which has attracted a lot of attention for several decades now, and it has been studied over the years in several different contexts. One advantage of considering non-linear electromagnetic Lagrangians is the fact that assuming appropriate non-linear electromagnetic sources, which in the weak field limit are reduced to the usual Maxwell's linear theory, one can generate a new class of Bardeen-like \cite{Bardeen,borde} BH solutions \cite{beato1,beato2,beato3,bronnikov,dymnikova,hayward,vagenas1,vagenas2} with certain desirable properties. In particular, those solutions, on the one hand, do have a horizon, which is the defining property of BHs, and, on the other hand, their curvature invariants, such as the Ricci or the Kretschmann scalar, are regular everywhere. This is to be contrasted to the standard Reissner-Nordstr{\"o}m solution \cite{RN}, which is characterized by a singularity at the origin. 
\\ 
Finally, Maxwell's theory may be easily generalized in a straightforward manner by a simple model called the Einstein-power-Maxwell (EpM) theory \cite{EpM1,EpM2,EpM3,EpM4,EpM5,EpM6,EpM7,EpM8,EpM9,EpM10,EpM11}. In this toy model, the theory is described by a Lagrangian density of the form $\mathcal{L}(F) \sim F^{\beta}$, where $\beta$ is an arbitrary rational number, $F \equiv F_{\mu \nu} F^{\mu \nu}$ is the Maxwell invariant, with 
$F_{\mu \nu} \equiv \partial_\mu A_\nu-\partial_\nu A_\mu$ 
being the field strength, and $A_\mu$ is the Maxwell potential. Even more, although our observable Universe seems to be four-dimensional, the question "How many dimensions are there?" is one of the fundamental questions that modern High Energy Physics tries to answer. Kaluza-Klein theories \cite{kaluza,klein}, Supergravity \cite{nilles} and Superstring/M-Theory \cite{ST1,ST2} have pushed forward the idea that extra spatial dimensions may exist. The advantage of the EpM theory is that it preserves the nice conformal properties of the four-dimensional Maxwell's theory in any number of space-time dimensionality $d$, provided the power $\beta$ is chosen to be $\beta=d/4$, as it is easy to verify that for this particular value the electromagnetic stress-energy tensor becomes traceless.
\\ 
In black hole physics, the impact of the SD scenario on properties of BHs has been studied over the last years, and it has been found that the scale dependence modifies the horizon, the thermodynamics, as well as the quasinormal spectra of classical BH backgrounds \cite{SD1,SD2,SD3,SD4,SD5,SD6,SD7,ourprd}. To the best of our knowledge, however, the impact of the SD scenario on four-dimensional charged black holes in the EpM theory has not been studied yet. In the present work, we propose to study for the first time the properties of scale-dependent BHs with a net electric charge in the EpM non-linear electrodynamics with a two-fold goal in mind. First, to fill a gap in the literature and, second, to make a direct comparison between i) SD BHs in the EpM theory versus their four-dimensional classical counterparts, and ii) the results obtained in the SD scenario versus those obtained by Renormalization Group (RG hereafter) improvement methods.

It is well-known that properties of physical systems depend on the dimensionality of space-time, see e.g. 
\cite{Kanti:2014vsa} for Hawking radiation from higher-dimensional black holes, and \cite{Konoplya:2004xx} for quasi-normal modes of black holes in Gauss-Bonnet gravity. Therefore, although EpM theory is better motivated in dimensions other than four, we still feel it would be interesting to see how the properties of black holes in this class of non-linear Electrodynamics change as we move from three to four dimensions.

Our work in the present article is organized as follows: In the next section, we review the classical EpM theory, while in section 3, we briefly present the BH solutions within EpM non-linear electrodynamics for arbitrary power $\beta$. In the fourth section, we comment on the null energy condition, an essential ingredient of the SD scenario. After that, in section 5, we discuss the properties of the SD charged BHs in four-dimensional EpM, and a comparison with the results produced by RG improvement methods is made as well. We finish our work with some concluding remarks in the last section.

\section{Classical Einstein-power-Maxwell theory}

This section is devoted to introducing a particular type of NLE that we are interested in. To be more precise, we will use the well--known EpM theory.
First, we remember that the standard Maxwell contribution to the action is defined as $F \equiv F^{\mu \nu} F_{\mu \nu}/4$. This solution has been extensively studied in the context of black holes in (2+1), (3+1), and in general in $d$ dimensional space-time. 
An apparently more complicated contribution to the action is obtained allowing powers of the aforementioned invariant, namely, now we accept a Lagrangian density of the form: $\mathcal{L}(F) \equiv D_1 F + D_2 F^2 + D_3 F^3 + \cdots + D_n F^n$. This Lagrangian density is not easy to investigate because, in general, it does not give exact solutions. A more treatable way to make progress is to take into account the special case $\mathcal{L}(F) = D^{\beta} |F|^{\beta}$, being $D$ a constant with appropriate units and $\beta$ a free dimensionless parameter. 

Those theories will then be investigated in the context of scale-dependent couplings. 
Thus, we will start considering the so--called EpM action without cosmological constant $(\Lambda_0 =0)$, 
assuming the aforementioned Lagrangian density, namely
\begin{align} \label{act1}
I_0[g_{\mu \nu}, A_{\mu}] &\equiv  \int {\mathrm {d}}^{4}x {\sqrt {-g}}\,
\bigg[\frac{1}{2\kappa_0} R - \frac{1}{e_{0}^{2\beta}}\mathcal{L}(F) \bigg],
\end{align}
where the parameters are defined as follow: $\kappa_0 \equiv 8\pi G_0$ is the gravitational coupling, $G_0$ is the dimensionful Newton's constant, 
$e_0$ is the dimensionful electromagnetic coupling constant, 
$R$ is the Ricci scalar and 
$F$ has the usual meaning. We use the metric signature $(-, +, +, +)$, and natural units ($c = \hbar = k_B = 1$) such that the action is dimensionless.
The power $\beta$ also appears in the exponent of the electromagnetic coupling. This inclusion is justified because we need to maintain the action dimensionless.
On the one hand, this generalized action also contains the classic Einstein-Maxwell case after the replacement $\beta =1$.
On the other hand, one can obtain deformed Maxwell solutions when $\beta \neq 1$.
In what follows, we shall consider the general case, namely, when $\beta$ is taken to be an arbitrary parameter. It is essential to point out that this generalization should recover the classical case, at least at a certain limit.
At this point we could consider a (naive) range $\beta \in \mathbb{R}^+$  \cite{Hassaine:2008pw}, but our solution could have restrictions on the values of the parameter $\beta$.

In the context of SD gravity in the presence of NLE, the corresponding equations of motion are
\begin{align}\label{Gmunu}
G_{\mu \nu} &= \frac{\kappa_0}{e_0^{2\beta}} T_{\mu \nu }.
\end{align}
The energy-momentum tensor $T_{\mu \nu}$ is associated to the electromagnetic field strength $F_{\mu \nu}$ through
\begin{align}\label{TNL}
T_{\mu \nu} &\equiv T_{\mu \nu}^{\text{EM}} = \ 
\mathcal{L}(F) g_{\mu \nu} - \mathcal{L}_{F} F_{\mu \gamma} F_{\nu}\ ^{\gamma},
\end{align}
remembering that $\mathcal{L}_F= \mathrm{d}\mathcal{L}/\mathrm{d}F$. Besides, for static spherically symmetric solutions the electric 
field $E(r)$ is given by 
\begin{align}\label{Fmunu}
F_{\mu \nu} &= (\delta_{\mu}^{r} \delta_{\nu}^{t} - \delta_{\nu}^{r} \delta_{\mu}^{t})E(r).
\end{align}
The variation of the classical action with respect to 
the field $A_{\mu}(x)$ gives simply
\begin{align} \label{decov}
D_{\mu}\left(\frac{\mathcal{L}_{F}F^{\mu\nu}}{e_0^{2\beta}}\right)& = 0,
\end{align}
where $e_0^{2\beta}$ is a constant.
Combining Eq. \eqref{Gmunu} with Eq. \eqref{decov} we are able to 
determine the set of functions $\{f_0(r), E_0(r)\}$.  
Also, it is important to point out that the classical version of this problem was certainly discussed before \cite{Hassaine:2008pw} as well as the corresponding SD Einstein-Maxwell case in (3+1) dimensions \cite{Rincon:2018dsq}. In both cases, the thermodynamic and the asymptotic properties were investigated in detail.

\section{Black hole solution for Einstein-Maxwell model of arbitrary power}
\label{Clasico}

The general metric ansatz assuming spherical symmetry is given by
\begin{equation}\label{metric}
\mathrm{d}s^2 = - f_0(r) \mathrm {d}t^2 + g_0(r) \mathrm {d}r^2 + r^2 \bigl(\mathrm {d}\theta^2 + \sin^2(\theta)\mathrm{d}\phi^2 \bigl),
\end{equation}
where $f_0(r)$ and $g_0(r)$ are the metric functions and can be linked via the Schwarzschild relation, i.e. $g_0(r)=f_0(r)^{-1}$. 
In GR, one only computes the electric field $E_0(r)$ and the lapse function $f_0(r)$ only by solving: i) the Einstein field equations and ii) the equation obtained after the variation of the classical action respect the potential $A_{\mu}$. For practical purposes, we will summarize the classical solutions for an arbitrary index, $\beta$.
Solving the Einstein field equations for the classical case we obtain:
\begin{align}
f_0(r) & = 1 + \frac{C}{r} + \frac{\tilde{B}}{r^{\alpha}},
\\
E_0(r) &=  \frac{\tilde{A}}{r^{\alpha }},
\end{align}
where the power $\alpha$ is linked to the power-Maxwell exponent as follows
\begin{align}
\alpha = \ & \frac{2}{2\beta-1},
\end{align}
and the case $\beta=1/2$ is excluded from the solution.
The pair $\{\tilde{A}, C\}$ are constants of integration directly related to the electric charge and the mass, respectively, of the black hole, while $\tilde{B}$ is given in terms of $\tilde{A}$ as follows \cite{Hassaine:2008pw}
\begin{align}
\tilde{B} &\equiv \kappa_0 \big( -2 \tilde{A} ^2 D\big)^{\beta} \frac{(1-2\beta)^2}{(2\beta -3)}.
\end{align}
Furthermore, $\tilde{A}$ is identified with the electric charge, $Q_0$, while $C=-2 G_0 M_0$ is proportional to the mass of the black hole, $M_0$. Throughout the manuscript, all classical quantities carry a sub-index "0", whereas scale-dependent quantities carry no sub-indices. Clearly, in the special case $\beta = 1$ and $\alpha=2$ the usual Reissner-Nordström (RN) black hole solution is recovered.

It is important to point out that $M_0$ is the ADM/Komar mass of the black hole. The ADM mass is defined via the ADM formalism at spatial infinity only. In practice, it is read off from the decay of $g_{00}$; the Komar mass is defined as a flux integral associated to the stationarity, and it can be computed at any closed 2-surface in a space-like surface. The Komar mass coincides with the ADM mass in the case of asymptotically flat space-times, such as the RN 
geometry.

Furthermore, we observe two natural ranges that respect the exponent $\alpha$, i.e., whether or not $1/r^{\alpha}$ goes faster to zero than the Schwarzschild potential $1/r$.
Regarding BH thermodynamics, we first should compute the classical horizon, $r_{0}, $which is obtained demanding that $f_0(r_0) = 0$. 
By writing the lapse function in terms of the horizons, we have
\begin{align}
f_0(r) &= \bigg[ 1 - \frac{r_A}{r} \bigg] 
\Bigg[
1 - 
\left[\frac{1 - \frac{r_A}{r_B}}{1 - \frac{r_A}{r}}\right] 
\Bigg[
\frac{\frac{r_A}{r} - \left(\frac{r_A}{r} \right)^{\alpha}}{\frac{r_A}{r_B} - \bigl(\frac{r_A}{r_B} \bigl)^{\alpha}}
\Bigg] 
\Bigg].
\end{align}
Firstly, notice that these horizons, $\{ r_A, r_B \}$,  can be linked to the BH mass and electric charge. Also, the classical BH horizon is the outer root of the lapse function, i.e., $r_0 \equiv \text{max}\{r_A,r_B\}$.
Given the non-trivial form of the lapse function, it is impossible to obtain the corresponding roots for an arbitrary index $ \alpha $, reason why we write it down implicitly.
Also, to get insights into this model, it is always useful to study some thermodynamic properties.
We can then define three quantities, {\it i. e.}, the Hawking temperature, $T_H$, the Bekenstein-Hawking entropy, $S$, and the specific heat, $C_Q$. Their 
corresponding expressions are given as follow
\begin{align}
T_0(r_0) &= \frac{1}{4 \pi} \Bigg|\lim_{r\rightarrow r_0} \frac{\partial_r g_{tt}}{\sqrt{-g_{tt}g_{rr}}} \Bigg|,
\\
S_0(r_0) &= \frac{\mathcal{A}_0}{4 G_0 },
\\
C_0(r_0) &= T \ \frac{\partial S}{\partial T} \ \Bigg{|}_{r_0},
\end{align}
being $\mathcal{A}_0$ the horizon area defined as
\begin{align}
\mathcal{A}_{0} &= \oint \mathrm{d^2}x \sqrt{h} = 4\pi r_0^2,
\end{align}
where $h_{ij}$ is the induced metric at the horizon, $r_0$.

\section{Scale dependent coupling and scale setting}
\label{scale_setting}

This section is devoted to summing up the main features and equations of motion for the SD EpM theory with an arbitrary index in four dimensions. 
The idea and spirit follow references  \cite{previous,Koch:2014joa,angel,Rincon:2017ypd,Rincon:2017ayr,Hernandez-Arboleda:2018qdo,Contreras:2017eza,Contreras:2018dhs,Contreras:2018swc,Rincon:2018lyd}.
and references therein.
\noindent In the SD formalism, the couplings evolve with the energy scale. In our case, we have two coupling functions to consider, {\it i. e.}, i) the Newton's coupling $G_k$ (which is related to the gravitational coupling by 
$\kappa_k \equiv 8 \pi G_k$), and ii) the electromagnetic coupling, $1/e_k$. 
Besides, there are three independent fields, which are the metric tensor, $g_{\mu \nu}(x)$, the electromagnetic four-potential, $A_{\mu}(x)$, and the scale field, $k(x)$, where $x^\mu \equiv x$ is any space-time point.
Notice that in Schwarzschild coordinates we can write: $x^\mu = \{t,r,\theta,\phi\}$, and due to the spherical symmetry all quantities depend on the radial coordinate only, e.g. $k(r)$.

The effective action for this theory takes the form
\begin{align}\label{Effective_Actiom}
\Gamma[g_{\mu \nu}, A_{\mu}, k] &=  \int {\mathrm {d}}^{4}x {\sqrt {-g}}\,
\bigg[\frac{1}{2\kappa_k} R - \frac{1}{e_{k}^{2\beta}}\mathcal{L}(F) \bigg].
\end{align}
Following the same strategy used in the non-improved case, we can obtain the equation of motion by taking the variation of (\ref{Effective_Actiom}) with respect to
$g_{\mu \nu}(x)$, namely
\begin{align}
G_{\mu\nu} &= \frac{\kappa_k}{e^{2\beta}_k}T^{\text{eff}}_{\mu\nu},
\end{align}
where the effective energy-momentum tensor is defined in such a way that the classical contribution is shifted by the inclusion of the Newton's scale-dependent coupling:
\begin{align}
T^{\text{eff}}_{\mu\nu} &= T^{\text{EM}}_{\mu\nu} - \frac{e^{2\beta}_k}{\kappa_k} \Delta t_{\mu \nu}.
\end{align}
To be more precise, $T^{\text{EM}}_{\mu\nu}$ is given by (\ref{TNL})
and the extra contribution $\Delta t_{\mu \nu}$ is 
\begin{align}
\Delta t_{\mu\nu} &= G_k \Bigl(g_{\mu \nu} \square - \nabla_{\mu} \nabla_{\nu}
\Bigl)G_k^{-1}.
\end{align}
Equivalently, the equations of motion for
the four-potential $A_{\mu}(x)$ when the electromagnetic coupling evolves take the following form
\begin{align} \label{decovcoupling}
D_{\mu}\left(\frac{\mathcal{L}_{F}F^{\mu\nu}}{e_k^{2\beta}}\right)& = 0.
\end{align}
At this level, some comments are in order. As we previously said, the SD scenario takes advantage of asymptotically safe gravity. In particular, in any quantum field theory, the renormalization scale $k$ has to be set to a quantity characterizing the physical system under consideration. Thus, for background solutions of the gap equations, the renormalization scale can evolve.
However, the price to pay is that the set of equations of motion does not close consistently. The latter means that the energy-momentum tensor could not be conserved for almost any choice of the functional dependence $k=k(x)$,
being $x$ an arbitrary coordinate. Also, an appropriated choice of $x$ depends of the symmetry and the problem itself, so, such identification will be specified later.
Also, such feature was also extensively investigated in the context
of renormalization group improvement of BHs in asymptotic safety scenarios \cite{Bonanno:2000ep,Bonanno:2006eu,Reuter:2006rg,Reuter:2010xb,Falls:2012nd,Cai:2010zh,Becker:2012js,Becker:2012jx,Koch:2013owa,Koch:2013rwa,Ward:2006vw,Burschil:2009va,Falls:2010he,Koch:2014cqa,Bonanno:2016dyv,Adeifeoba:2018ydh,Platania:2019kyx}.
The loss of conservation laws comes from the fact that there is one consistency equation missing.
%

Varying the effective action (\ref{Effective_Actiom}) with respect to the scale field $k(x)$ we can find the missing equation, i.e.
\begin{equation}\label{vary}
\frac{\mathrm{d}}{\mathrm{d} k} \Gamma[g_{\mu \nu}, A_{\mu}, k]=0,
\end{equation}
which can be considered as as variational scale setting procedure \cite{Reuter:2003ca,Koch:2010nn,Domazet:2012tw,Koch:2014joa,Contreras:2016mdt}.
To achieve the conservation of the stress-energy tensor, we can combine Eq.~\eqref{vary} with the above equations of motion. Further details regarding the split symmetry within the functional renormalization group equations support this approach of dynamic scale setting can be found in \cite{Percacci:2016arh}.
Now, it should be noticed that the implementation of the variational procedure (\ref{vary}) requires the knowledge of the corresponding $\beta$-functions of the problem. This, however, is a strong disadvantage because those are not unique. Thus, to by-pass such a problem, we will close our system by adding a constraint on one energy condition. The latter has also been used in some concrete BH solutions
\cite{Contreras:2013hua,Koch:2015nva,angel,Rincon:2017ypd,Contreras:2017eza,
Contreras:2018dhs,Rincon:2018sgd}.
We then adopt the same route by imposing the so-called null energy condition (NEC) to study EpM BHs in four dimensions.
%

\section{The null energy condition}\label{NEC}

To close our system, we need to select a supplementary condition. As we have treated in previous works, we will take advantage of the so-called NEC. 
In general, an energy condition is an extra relation that we impose on the energy-momentum tensor to try to capture the idea that energy should not be negative~\cite{Curiel:2014zba}. We usually have four energy conditions: i) dominant, ii) weak, iii) strong, and finally, iv) the NEC. In a well-defined problem in GR, such restrictions are satisfied, although in some other cases, they can be violated \cite{Rubakov:2014jja,Wald:1984rg}.

It is also remarkable that the NEC is a critical ingredient in the  Penrose singularity theorem \cite{Penrose:1964wq}. Given that our solutions belong to an extension of GR, it is also suitable to maintain at least one energy condition. In such a sense, the NEC is the less restrictive of them.
Assuming as valid the classical NEC, we always have a singularity. Thus, any contracting Universe ends up in a singularity, provided its spatial curvature is dynamically negligible \cite{Rubakov:2014jja}. NECs can  even be extended to quantum formulations~\cite{Grumiller:2019xna,Ecker:2019ocp}.
Given the relevance and particular beauty of the NEC, we will focus our attention on it. Our starting point is to consider certain null vector, called $\ell^{\mu}$, and to contract it with the matter stress energy tensor as the NEC demands, {\it i. e.} : 
\begin{align}
T^{m}_{\mu \nu} \ell^{\mu} \ell^{\nu} \geq 0.
\end{align}
The latter strategy was also used in Ref. \cite{angel} inspired by  Jacobson's idea \cite{Jacobson:2007tj} on getting acceptable physical solutions. Besides, the relevance of the NEC becomes notorious when we recognize that such a condition is not optional in proving some fundamental BH theorems, such as the no-hair theorem \cite{Heusler:1996ft}, and the second law of black hole thermodynamics \cite{Bardeen:1973gs}.
In the SD scenario we maintain the same condition in a more restrictive and thus more useful form
by making the inequality an equality 
\begin{align}\label{NEC1}
T^{\text{eff}}_{\mu \nu} \ell^{\mu} \ell^{\nu} = \bigg(T^{\text{EM}}_{\mu \nu} - \frac{e^{2\beta}_k}{\kappa_k} \Delta t_{\mu \nu}\bigg) \ell^{\mu} \ell^{\nu} = 0.
\end{align}
For the null vector we choose a
radial null vector $\ell^{\mu}=\{f^{-1/2}, f^{1/2}, 0\}$.
Since the electromagnetic contribution to the effective stress energy tensor (\ref{TNL})
satisfies the NEC (\ref{NEC1}) by construction, the same has to hold
for the additional contribution introduced due to the SD of the gravitational coupling: 
\begin{align}\label{Delta_t}
\Delta t_{\mu \nu} \ell^{\mu} \ell^{\nu} = 0.
\end{align}
%
%

\section{Scale dependent Einstein-power-Maxwell theory}
\label{solution}

\subsection{Solution}
Now, we will compute the solutions for our SD system of differential equations. In classical BH solutions, we need to find the lapse function and the electric field. Instead, we need to compute the same functions in the SD BH version and the corresponding Newton and electromagnetic couplings. 
Thus, we can sum up it as follows:
\begin{align}
\{f_0(r), E_0(r), G_0, e_0\} \rightarrow \{f(r), E(r), G(r), e(r)\}
\end{align}
Notice that in Schwarzschild coordinates we can write: $x^\mu = \{t,r,\theta,\phi\}$, and then due to spherical symmetry $x^{\mu} = r$ only.
The first step to find the full solution is to calculate Newton's coupling. We first solve it because, in practice, when the NEC is used, the differential equation for $G(r)$ is not coupled to the rest of the functions. We can then solve it directly. The differential equation is written as
\begin{align}
G(r)\frac{{\mathrm {d}}^2 G(r)}{\mathrm {d} r^2} - 2 \left(\frac{\mathrm {d} G(r)}{\mathrm {d} r}\right)^2=0,
\end{align}
which allows us to obtain 
\begin{align}\label{Gr}
G(r) = \frac{G_0}{1 + \epsilon r},
\end{align}
after a suitable choice of integration constants. 
Be aware and notice that the constant $\epsilon$ encodes quantum features; therefore, the classical solution is recovered when $\epsilon$ goes to zero.
The second step is to solve the equation of motion for the 4-potential given by Eq. \eqref{decovcoupling}.
\begin{align}
\frac{\mathrm{d} E(r)}{\mathrm{d}r} - 
\Bigg[
\bigg(1 + \frac{1}{2}\alpha \bigg)
\frac{ e'(r)}{e(r)}-\frac{\alpha }{r}
\Bigg]
E(r) &=0.
\end{align}
In light of the radial dependency of the electric field, the differential equation can also be solved directly to get
\begin{align}
E(r) = \ & \tilde{A} \Bigg[\frac{e(r)^{\left(1 + \frac{1}{2}\alpha \right)}}{r^{\alpha}}\Bigg].
\end{align}
At this level, we would like to clarify the meaning of some of the integration constants. Firstly, we introduce the so-called running parameter, $\epsilon$, which controls the strength of the scale dependence; $G_0$ is the classical Newton's coupling constant and, finally, $\tilde{A}$ is a coupling constant which controls the strength of the electric field.
%
%
%
%
%
\begin{figure*}[ht]
\centering
\includegraphics[width=0.48\textwidth]{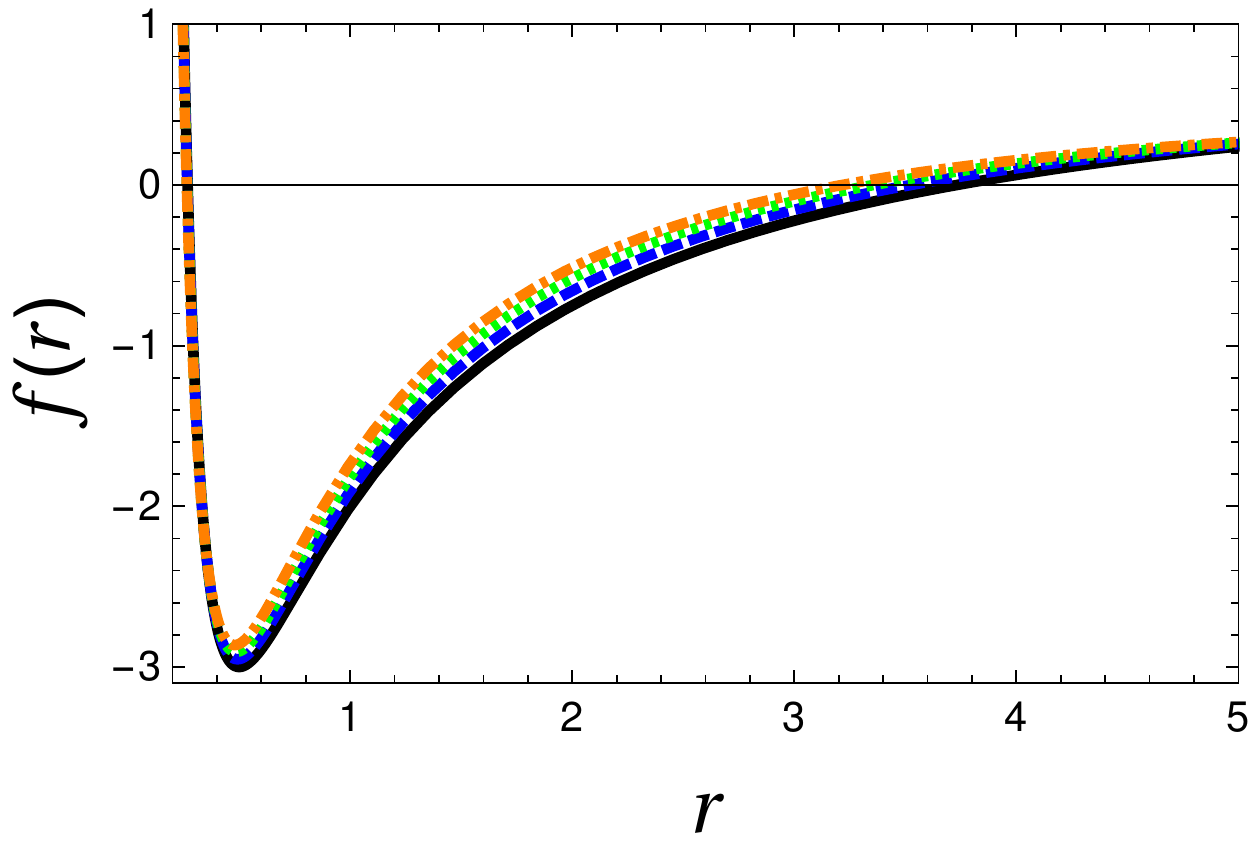}   \
\includegraphics[width=0.48\textwidth]{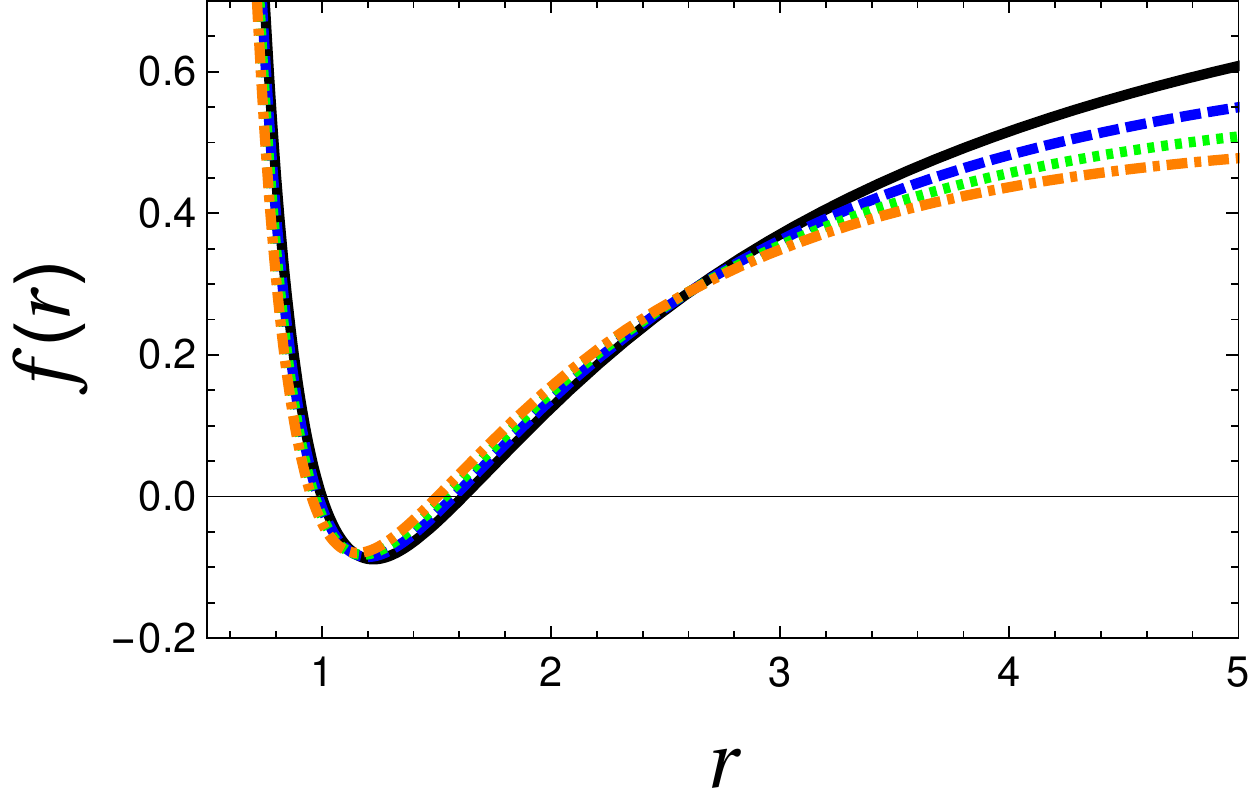} 
\\
\includegraphics[width=0.48\textwidth]{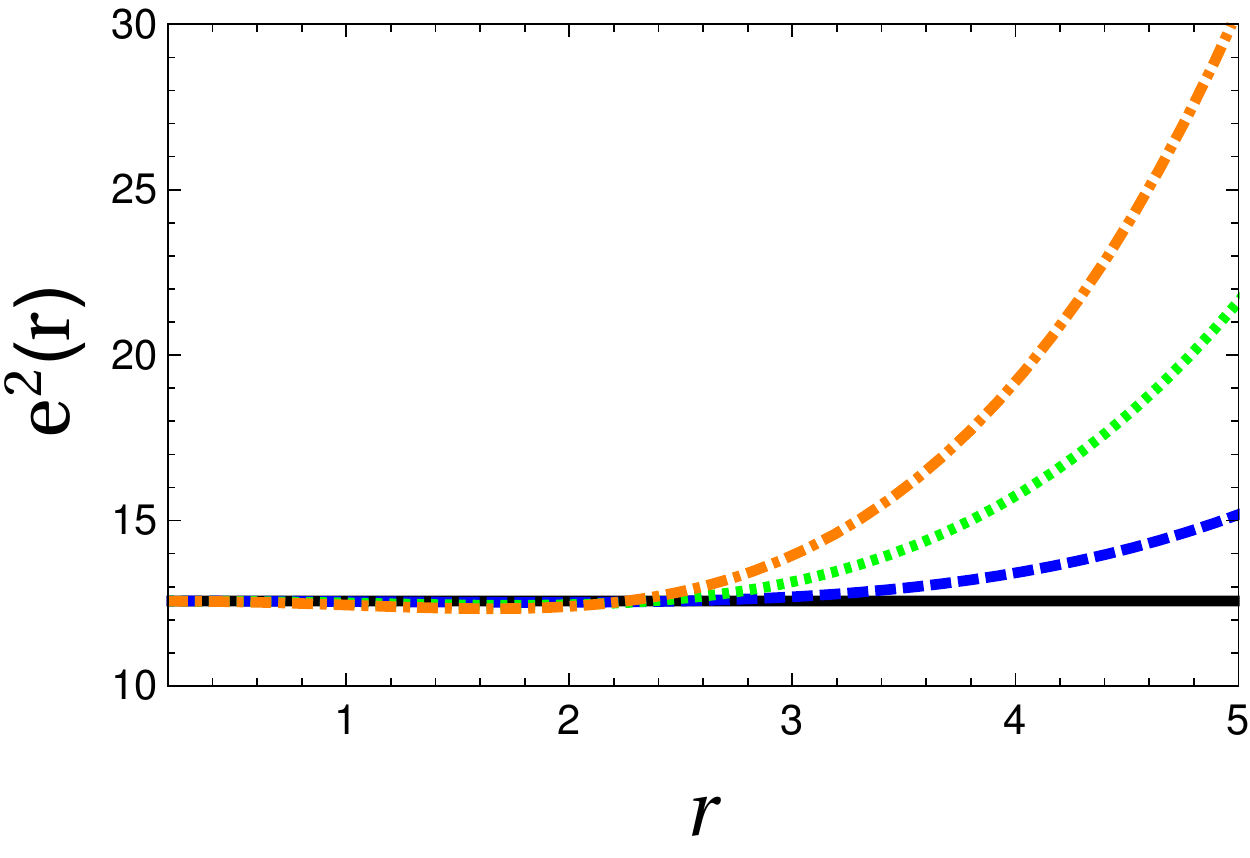}   \
\includegraphics[width=0.48\textwidth]{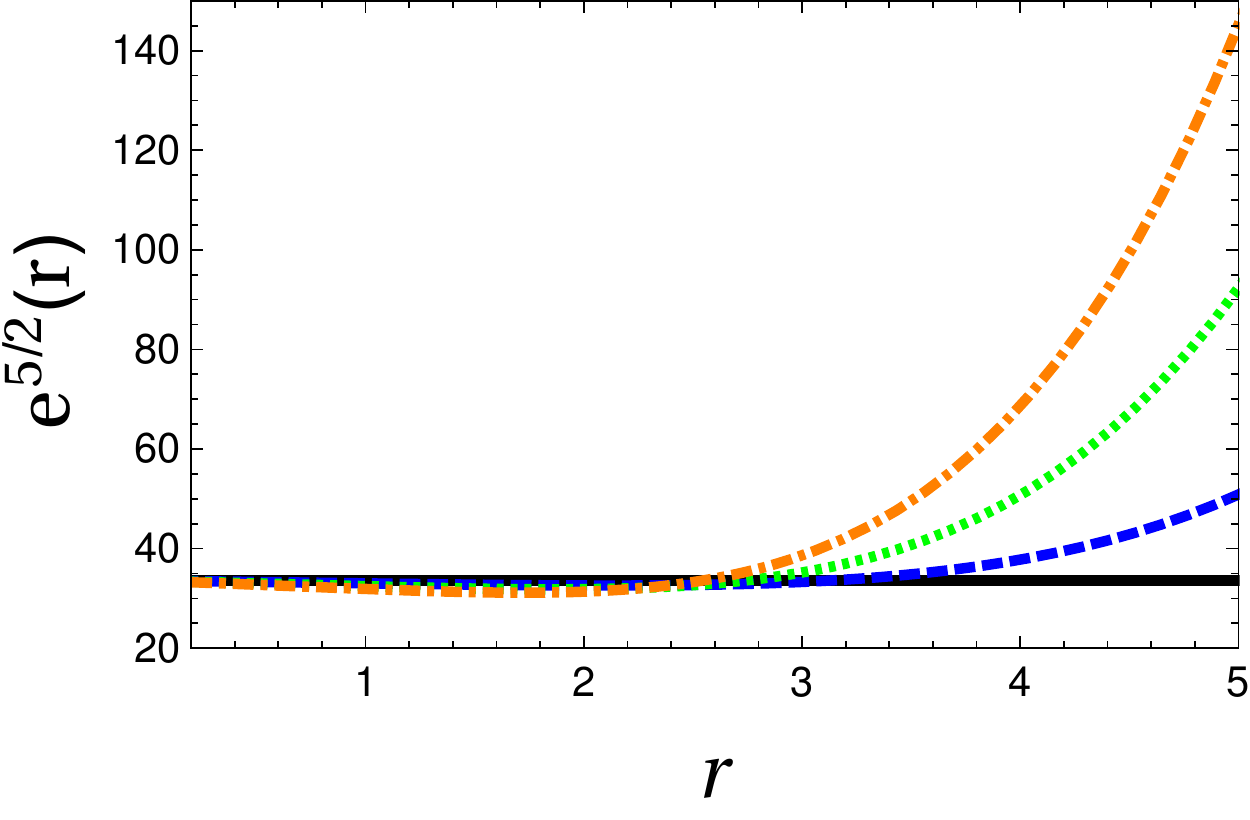} \
\caption{
The lapse function $f(r)$ and the electromagnetic coupling $e(r)^{1 + \frac{1}{2}\alpha}$ versus radial coordinate $r$ for  two concrete cases. The first 
line correspond to the lapse function while the second line correspond to the electromagnetic coupling. The first (left),  
and second (right) column correspond to the cases $\alpha= \{2,3\}$ respectively. We show the classical model (solid black line) and three different cases for each figure: 
i) $\epsilon = 0.033$ (dashed blue line), ii) $\epsilon = 0.067$  (dotted red line) and iii) for $\epsilon = 0.010$  (dotted dashed green line). 
We have used the set 
$\{Q_0, e_0, G_0, M_0\} = \{ 1, 1/(2\sqrt{\pi}), 1/(16 \pi^2), 32 \pi^2 \}$ . 
The numerical values of the event horizon are: 
i) for $\alpha=2$, 
$\{r_H(\epsilon=0), r_H(\epsilon=0.033), r_H(\epsilon=0.067), r_H(\epsilon=0.010) \} = \{3.73204, 3.52496, 3.35651, 3.21518 \}$ 
and
ii) for $\alpha=3$,
$\{r_H(\epsilon=0), r_H(\epsilon=0.033), r_H(\epsilon=0.067), r_H(\epsilon=0.010) \} = \{1.61803, 1.57659, 1.539, 1.50465 \}$.
%
%
}
\label{fig:1}
\end{figure*}
Solving for the remaining function $f(r)$ and $e(r)$ we obtain the non-trivial solutions 
\begin{align}\label{fr}
\begin{split}
f(r) &= \frac{1}{6} (1 + \epsilon r )^{-(1 + \frac{1}{2}\alpha)}
\Bigg[
\frac{6 \tilde{B}}{r^{\alpha }} - \frac{(-\epsilon r )^{-\alpha }}{3 \alpha +2} \times
\\
& \times
	\Bigg\{
	   \bigg(4 \alpha  (\alpha +1)+\epsilon  (6 (3 \alpha +2) C   
	   \\
		&	 +  
	   (3 \alpha -2) (7 \alpha +4) r)+3 \alpha  (3 \alpha -2) r^2 \epsilon ^2
		\bigg)
\\		
& \times		
	B_{-r \epsilon }\left(\alpha -1, 1 + \frac{1}{2}\alpha \right)
	- \alpha  (1 + \epsilon r ) \times
\\		
& \times
	\bigg( \alpha  (3 r \epsilon +4)+4 \bigg) B_{-r \epsilon }\left(\alpha -1,\frac{1}{2}\alpha\right)
	\Bigg\}
\Bigg],
\end{split}
\end{align}
\begin{align}\label{er}
\begin{split}
e(r)^{\left(1 + \frac{1}{2}\alpha\right)} 
&=
W_0
\frac{(1 + \epsilon r)^{-\frac{\alpha }{2}-2}}{r} \times
\\
& \times
\Bigg[
W_0(r,\epsilon) 
+ 
\frac{r^{\alpha +1} (-r \epsilon )^{-\alpha }}{3 \alpha +2} 
\times
\\
& \times	
	\Bigg\{
	W_1(r,\epsilon) B_{-r \epsilon }\left(\alpha -1, 1 + \frac{1}{2}\alpha \right) 
\\
& + \ \ W_2(r,\epsilon) B_{-r \epsilon }\left(\alpha -1,\frac{1}{2}\alpha\right)
	\Bigg\}
\Bigg],
\end{split}
\end{align}
where we have defined the intermediate functions as
\begin{align}
W_0 &\equiv \frac{2^{\frac{1}{\alpha }-\frac{13}{2}} \alpha}{3 \pi G_0}  
\left(
\tilde{A}D^{\frac{1}{2}}
\right)^{-
\left(
\frac{\alpha + 2}{\alpha}
\right)
} ,
\end{align}
\begin{align}
W_0(r,\epsilon) &\equiv 6  \tilde{B} r
 \left(
 \alpha  (3 r \epsilon +2)^2 - 2 (3 r \epsilon  (r \epsilon +2)+2)
 \right)
  -
\nonumber
\\
& 
\ \ \ \ 
2 r^{\alpha } (r \epsilon +1)^{\frac{\alpha }{2}+1} 
(
6 C (3 r \epsilon +2) + 
\\
&
\ \ \ \ 
r 
(
3 \alpha  (r \epsilon +2) (3 r \epsilon +2) - 4 (3 r \epsilon  (r \epsilon +3) + 5)  
)
) ,
\nonumber
\end{align}
\begin{align}
W_1(r,\epsilon) &\equiv 4 \alpha  (3 r \epsilon +1) 
\left(
6 \epsilon  (C+r) - 3 r^3 \epsilon ^3 + 4 
\right) + 
\nonumber
\\
&
\ \ \ \
2 \alpha ^2 \epsilon  
(2 r (9 r \epsilon  (3 r \epsilon  (r \epsilon +4)+14)+41) - 
\\
&
\ \ \ \ 
9 C (3 r \epsilon +2)^2) ) + 8 \epsilon  (3 C (3 r \epsilon  (r \epsilon +2)+2)-
\nonumber
\\
&
\ \ \ \ 
2 r (2 r \epsilon +1) (3 r \epsilon +4)) - \alpha ^3 (3 r \epsilon +2)^2 \times
\nonumber
\\
&
\ \ \ \
(3 r \epsilon  (3 r \epsilon +7)+4) ,
\nonumber
\\
W_2(r,\epsilon) &\equiv 
\alpha  (r \epsilon +1) 
(
\alpha ^2 (3 r \epsilon +4) (3 r \epsilon +2)^2 - 6 \alpha  r \epsilon
\\
&
\ \ \ \ 
\times (r \epsilon  (3 r \epsilon +10)+6)-16 (r \epsilon +1) (3 r \epsilon +1)
) .
\nonumber
\end{align}
The solution is expressed in terms of the incomplete Beta function $B_{z}(a,b)$ defined as
\begin{align}
B_{z}(a,b) \equiv \int_{0}^{z} d \tau \  {\tau}^{a-1} (1-\tau)^{b-1} .
\end{align}
In particular, notice that $B_{z}(a,b)$ has a branch cut discontinuity in the complex $z$ plane running from $-\infty$ to $0$. 
Similarly $B_{z}(a,b)$ can also be defined as
follows:
\begin{align}
B_{z}(a,b) = \ & z^a \sum _{n=0}^{\infty } \frac{(1-b)_n}{n! (a+n)} z^n,
\end{align}
where $(\cdots)_n$ is the Pochhammer symbol, i.e.
\begin{equation}
(\cdots)_{n} = \frac{\Gamma(x + n)}{\Gamma(x)} .
\end{equation}
An interesting case appears when $z=1$. In such circumstance, the incomplete beta function $B_{z}(a,b)$ becomes to the usual beta function $B(a,b)$.
%

\subsection{Setting the integration constants}
Up to now, our generalized SD solution has a few arbitrary constants. To obtain them, we usually demand that the new solution should converge to the classical one when $\epsilon$ is taken to be zero. 
We must emphasize the number of integration constants involved in the problem. Analogously to the lower-dimensional case, the SD gravitational coupling introduces two of them. In this case, they are the classical Newton's constant $G_0$ and the running parameter $\epsilon$. Also, the electromagnetic field shows an additional integration constant $\tilde{A}$, and the solution of the lapse function is parametrized by two additional integration constants $\tilde{B}$ and $ C $. As was pointed out above, $\tilde{B}$ and $ C $ are related to the electric charge and classical mass, respectively, and the value $\tilde{A}$ is connected to the conventional charge.
Taking the limit when $\epsilon \rightarrow 0$ we have
\begin{align}
\lim_{\epsilon \rightarrow 0} f(r) &= f_0(r) = 1 + \frac{C/(\alpha -1) }{r}+ \frac{\tilde{B}}{r^{\alpha }},
\nonumber
\\
\lim_{\epsilon \rightarrow 0} E(r) &= E_0(r) = \tilde{A} \Bigg[\frac{\xi^{1+ \frac{1}{2}\alpha}}{r^{\alpha }}\Bigg] ,
\nonumber
\\
\lim_{\epsilon \rightarrow 0} G(r) &= G_0, 
\\
\lim_{\epsilon \rightarrow 0} e(r)^{1 + \frac{1}{2}\alpha } &= 
\xi^{1 + \frac{1}{2}\alpha } \equiv 
\frac{\tilde{B} \alpha(\alpha-1) }{16 \pi G_0} 
\Bigg[\tilde{A} \left[\frac{D}{2} \right]^{\frac{1}{2}}
\Bigg]^{-\frac{\alpha +2}{\alpha }} .
\nonumber
\end{align}
Also, for Maxwell electrodynamics ($\alpha = 2$) we recover the well-known family of solutions, i.e., 
\begin{align}
\lim_{\epsilon \rightarrow 0} f(r) &= f_0(r) = 1 + \frac{C}{r}+ \frac{\tilde{B}}{r^{2}},
\nonumber
\\
\lim_{\epsilon \rightarrow 0} E(r) &= E_0(r) = \tilde{A} \Bigg[ \ \frac{\xi^{2}}{r^{2}} \ \Bigg] ,
\nonumber
\\
\lim_{\epsilon \rightarrow 0} G(r) &= G_0, 
\\
\lim_{\epsilon \rightarrow 0} e(r)^{2} &= 
\xi^{2} = \frac{\tilde{B}}{4 \pi G_0  \tilde{A}^2 D } .
\nonumber
\end{align}
We can now take advantage of the SD Maxwell BH solution previously obtained in \cite{Koch:2015nva}. Following that setting, we take the parameters $\{\tilde{A}, \tilde{B},C,D\}$ as follow
\begin{align}
\tilde{A} &= \frac {Q_0^2}{4 \pi e_ 0^2},\\
\tilde{B} &= \frac {4 \pi G_0} {e_0^2} Q_0^2,\\
C &= -2 G_0 M_0,\\
D &= \frac{1}{Q_0^2}.
\end{align}
Thus, we will finally have the set $\{ G_0, e_0, Q_0, M_0, \epsilon \} $. Also, we can observe how the running of the gravitational coupling distorts the solution. 

To get some insights, we expand the solution for small values of the running parameter $\epsilon$ up to second-order to obtain to get
\begin{align}
\begin{split}
f(r) &\approx f_0(r) + \frac{\left(\alpha ^2-1\right) r \epsilon  ((3 \alpha +4) r \epsilon -2 (\alpha +2))}{2 (\alpha -1) (\alpha +1) (\alpha +2)} 
\\
&
+
\frac{1}{8} \tilde{B} r^{-\alpha } ((\alpha +2) r \epsilon  ((\alpha +4) r \epsilon -4)+8)
\\
&
+
\frac{C \left(4 (\alpha +1)+3 (3 \alpha +2) r^2 \epsilon ^2-6 (\alpha +1) r \epsilon \right)}{4 (\alpha -1) (\alpha +1) r} ,
\end{split}
\end{align}
and 
\begin{align}
e(r)^{1+\frac{1}{2}\alpha} \approx 
\frac{\xi^{1+ \frac{1}{2}\alpha}}{24 (\alpha -1) \tilde{B} r}
\Bigg[
\frac{2 Z r^{\alpha }}{2 + \alpha - 2 \alpha^2 - \alpha^3}+Y
\Bigg] ,
\end{align}
where $Y\equiv Y(r,\epsilon)$ and $Z \equiv Z(r, \epsilon)$ are supplementary functions defined as
\begin{align}
Z(r,\epsilon) &=  6 (\alpha +2) C 
\bigl(
\alpha ^2 (r \epsilon -2) (r \epsilon +1)+r \epsilon  (1-4 r \epsilon )+2
\bigl)
\nonumber
\\
&
\ \ \ +
\bigl(
\alpha ^2-1
\bigl) r 
\bigl(-4 (\alpha +2) (3 \alpha -5) + (\alpha  (3 \alpha +2) 
\nonumber
\\
&
\ \ \ - 44) r^2 \epsilon ^2-4 (\alpha +2) (3 \alpha -4) r \epsilon 
\bigl) ,
\\
Y(r,\epsilon) &= 3 \tilde{B} r 
\bigl( 
8 (\alpha -1)+(\alpha -3) (\alpha -2) (\alpha +2) r^2 \epsilon ^2 - 
\nonumber
\\
&
\ \ \ \ 4 (\alpha -2) (\alpha -1) r \epsilon 
\bigl) 
+ 
2 r^{\alpha } 
\bigl( 
6 C (r \epsilon -2) (r \epsilon +1) + 
\nonumber
\\
&
\ \ \ \ (3 \alpha -4) r^2 \epsilon  (r \epsilon -4)+4 (5-3 \alpha ) r
\bigl) .
\end{align}
Finally, the electric field and Newton's coupling take the form
\begin{align}
E(r) &\approx 
\frac{\tilde{A} \xi^{1+ \frac{1}{2}\alpha}}{24 (\alpha -1) \tilde{B} r^{1+\alpha}}
\Bigg[
\frac{2 Z r^{\alpha }}{2 + \alpha - 2 \alpha^2 - \alpha^3}+Y
\Bigg] ,
\\
G(r) &\approx G_0 
\left( 
1 - (\epsilon r) + (\epsilon r)^2
\right) .
\end{align}
It is essential to point out that certain values of the power $\alpha$ are not allowed. We read-off those values from the corresponding lapse function. Firstly, $\alpha = 1$ is excluded from the solution to satisfy the classical solution. After that, the SD lapse function has other problematic points to be excluded from the general solution. Thus, $\alpha = -1$ and $\alpha = -2$ are ignored too. 
To show the lapse function's behavior, we will take benchmarks to get insights about the underlying physics. Fig \eqref{fig:1} can be observed the behavior of the lapse function and the electromagnetic coupling for different values of the parameter $ \alpha $.

\section{Invariants}
In this part, we will briefly summarize how the Ricci scalar looks like when Newton's coupling constant becomes SD in light of an EpM source. The Kretschmann scalar is optional in this case, the reason why we will omit such a computation.
Thus, the Ricci scalar take the complicated form
\begin{align}
R &=
\frac{1}{24 (3 \alpha +2)}
\Bigg[
r^{-\alpha -3} (-r \epsilon )^{-\alpha } (r \epsilon +1)^{-\frac{\alpha }{2}-3}
\nonumber
\\
& \times 
(2 (-r \epsilon )^{\alpha } 
( r^{\alpha } (r \epsilon +1)^{\frac{\alpha }{2}+1} 
(6 (3 \alpha +2) C (2 \alpha +3 \alpha  r \epsilon -4)
\nonumber
\\
& 
+ 
r (4 (\alpha -2) (\alpha  (9 \alpha -17)-14) + 3 
(\alpha  
(9 \alpha ^2 - 30 \alpha +40)
\nonumber
\\
& 
+ 
16) r^2 \epsilon ^2 + 8 (\alpha  (3 \alpha  (3 \alpha -8)+25)+16) r \epsilon ))
\nonumber
\\
& 
- 
3 (3 \alpha +2) \tilde{B} r 
(
\alpha ^2 (3 r \epsilon +2)^2 - 6 \alpha  (r \epsilon  (r \epsilon +4)+2)+8 ) )
\nonumber
\\
& +
r^{\alpha +1} ( ( 16 (\alpha -2) (\alpha -1) \alpha  (\alpha +1) + (3 \alpha -2) r^2 \epsilon ^3 
\nonumber
\\
& 
\times
(18 \alpha  (3 \alpha +2) C + (3 \alpha -4) (\alpha  (33 \alpha -50)-8) r) 
\\
& 
+ 
4 r \epsilon ^2 (18 (\alpha -2) \alpha  (3 \alpha +2) C + (3 \alpha -2) (\alpha  (\alpha  (27 \alpha
\nonumber
\\
& 
-
83)+34)+24) r) + 4 (\alpha -2) \epsilon  (6 (\alpha -1) (3 \alpha +2) C
\nonumber
\\
& 
+ 
(\alpha  (\alpha  (33 \alpha -53)-2)+24) r) + 9 (\alpha -2) \alpha  (3 \alpha -4) 
\nonumber
\\
& 
\times
(3 \alpha -2) r^4 \epsilon^4 ) 
B_{-r \epsilon }
\Bigl(
\alpha -1, \frac{\alpha }{2} + 1 
\Bigl) - \alpha  (r \epsilon +1)
\nonumber
\\
& 
\times 
(\alpha ^3 (3 r \epsilon +2)^2 (3 r \epsilon +4) - 2 \alpha ^2 (r \epsilon  (3 r \epsilon  (15 r \epsilon +32)
\nonumber
\\
& 
+
70) + 16) + 8 \alpha  (r \epsilon  (r \epsilon  (9 r \epsilon +13) + 1) - 2) 
\nonumber
\\
& 
+ 
32 (r \epsilon +1)^2) 
B_{-r \epsilon }
\Bigl(\alpha -1,\frac{\alpha }{2} \Bigl)
)
)
\Bigg] .
\nonumber
\end{align}
Also, notice that, different from the classical case, 
$\alpha = -2/3$ makes that the Ricci scalar blows up. Besides, as the incomplete beta functions are present, negative values of $ \alpha $ should be ignored to maintain a non-singular Ricci scalar. The classical case can be recovered demanding that $\epsilon \rightarrow 0$ to get
\begin{align}
R_0 \equiv \lim_{\epsilon \rightarrow 0} R = -\frac{\tilde{B}}{r^{\alpha + 2}} (\alpha - 2) (\alpha -1) .
\end{align}
Finally, the Einstein-Maxwell solutions are obtained when $\alpha=2$ to achieve a null scalar.

\section{Horizon and Thermodynamics}
This section is dedicated to reviewing the main thermodynamic properties in the SD scenario. 
Some details to the computation of the corresponding thermodynamics properties can be consulted,for instance, in \cite{delaCruzDombriz:2009et}.
\subsection{Black hole horizon}
An essential ingredient to analyze the correspondent thermodynamics is the BH horizon. Thus, the horizon where the temperature, entropy, and capacity heat are evaluated is precisely the reason why it will compute it.
The BH horizon is found by demanding that $f(r_H) = 0$. In the simplest cases, we might obtain the explicit form of $r_H$. In this case, however, such a task is not possible to achieve. That problem is indeed present in the classical solution for an arbitrary EpM index.
Even though, when the arbitrary index is fixed, we could then obtain a tractable lapse function and an analytical expression for the horizon. 
Now, in the SD solution, we will take two concrete cases: $\alpha= \{2, 3\}$, to show their behavior.
As can be observed in Fig.~\eqref{fig:2} (left panel) the BH horizon decreases when the running parameter increases. The latter can be confirmed by taking, for instance, $\alpha=2$ and the running parameter close to zero, namely:
\begin{align}
r_H(\alpha=2) &= r_0 \Bigg[1 - \frac{1}{2} \epsilon r_0 + \mathcal{O}(\epsilon^2) \Bigg] ,
\end{align}
where $r_0$ is the classical BH horizon. We then confirm that the SD BH horizon is smaller than its classical counterpart, its major deviation appearing when $\epsilon r_0$ is significant compared to unity. 
In Fig. \eqref{fig:1} we have added the numerical values of the event horizon for reference. We have also checked, and a second-order expansion in $\epsilon$ is enough to make the exact value and the approximated value quite close. 
Be aware and notice that, for a given value of $\alpha$, the classical BH mass should be related to the other parameters of the theory. To be more precise, it should satisfy:
\begin{equation} \label{arbitrary}
M \geq
\left\{
\begin{array}{lcl}
\left(\frac{4\pi Q_0^2}{e_0^2 G_0}\right)^{1/2} & \mbox{ if } & \alpha = 2  , 
\\
\left(\frac{27 \pi Q_0^2}{e_0^2 G_0^2}\right)^{1/3}  & \mbox{ if } & \alpha = 3 ,
\end{array}
\right.
\end{equation}
for the classical solution.
The latter bounds are obtained to bypass troubles due to the appearance of naked singularities in the classical solution.
It should also be pointed out that the classical black hole mass is not obtained using the counter-term method or the Hamiltonian approach. Instead, such a parameter is found under the appropriate identification with the classical black hole solution.
Keep in mind that the SD BH solution includes the classical solution. The latter means that such a minimum value of the mass should be considered as a naive restriction, still in the SD case. It is remarkable that, at least for $\alpha=2$ in the SD scenario, the minimum bound of the BH mass is maintained, i.e., $M_\text{min}$ does not depend on $\epsilon$. For
$\alpha=3$ the situation is blurred due to the special functions in which the lapse function is written. 
The latter can be understood as follows: considering that the event horizon $r_H$ (which should be higher than zero) is used to read-off some bound for $M_0$, $r_H$ should be relatively simple. However, for $\alpha=3$, the black hole horizon is not analytical; therefore,  it is impossible to collect $M_0$ and build the corresponding bound (as was made for the classical case). Thus, for the SD case, the mass bound for $\alpha=3$ is impossible to obtain.
\subsection{Temperature}
In theories beyond Einstein's gravity, it is still possible to obtain the Hawking temperature following the usual route.
The starting point is the Euclidean action method \cite{Cai:1994np}. First, note that the metric can be written in terms of the Euclidean time $\tilde{\tau}$ after the change $t \rightarrow -i \tilde{\tau}$
\begin{equation}\label{metric2}
\mathrm{d}s^2 = f_0(r) \mathrm {d}\tilde{\tau}^2 + g_0(r) \mathrm {d}r^2 + r^2 \bigl(\mathrm {d}\theta^2 + \sin^2(\theta)\mathrm{d}\phi^2 \bigl),
\end{equation}
we then consider the requirement of the absence of the conical singularity in the Euclidean space-time \eqref{metric2} causes the Euclidean time $\tilde{\tau}$ to have a period $\beta_0$, which varies that the temperature is given by

\begin{align}
T_H &= \frac{1}{4 \pi} \Bigg|\lim_{r\rightarrow r_H} \frac{\partial_r g_{tt}}{\sqrt{-g_{tt}g_{rr}}} \Bigg|.
\end{align}

The complexity of the metric potentials makes it impossible to obtain an explicit form of the temperature. However, implicitly, writing the event horizon and taking advantage of small values of the running parameter might bypass the problem, at least for concrete cases. In Fig. \eqref{fig:2} (middle panel) we show the behaviour of the Hawking temperature for $\alpha = \{2, 3\}$. Notice that, starting from a certain point, the Hawking temperature increases to reach a maximum value, and, after that, it decreases when the mass $ M \equiv M_0$ increases. Also, a small, but still the noticeable difference is appreciated when the running parameter is modified. Thus, when the running parameter is turned on, the Hawking temperature increases with respect to its classical value.
To show the impact of SD gravity on classical solutions, we take the simplest case (i.e., Einstein-Maxwell). We observe that the correction to the background solution appears at second order in $\epsilon$ and also increases the effective temperature. 
\begin{align}
T_H(\alpha = 2) &\equiv T_0(\alpha=2) \Bigg| 1 + \frac{1}{4} \bigl( \epsilon r_0 \bigl)^2 \ + \ \mathcal{O}(\epsilon^3)    \Bigg| .
\end{align}
Finally, the case $\alpha=3$ is investigated only numerically due to the complicated expression for the lapse function. 
\subsection{Entropy and Heat capacity}
\begin{figure*}[ht]
\centering
\includegraphics[width=0.32\textwidth]{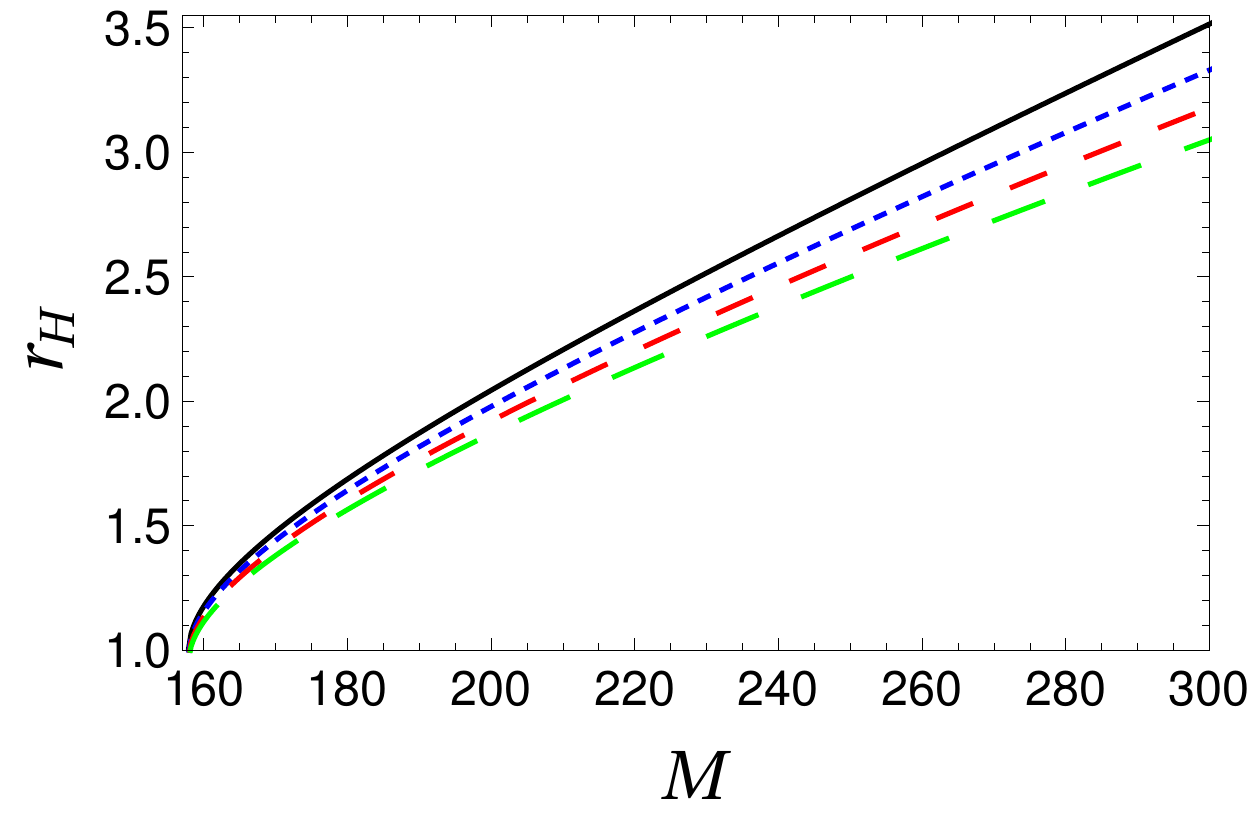}   \
\includegraphics[width=0.32\textwidth]{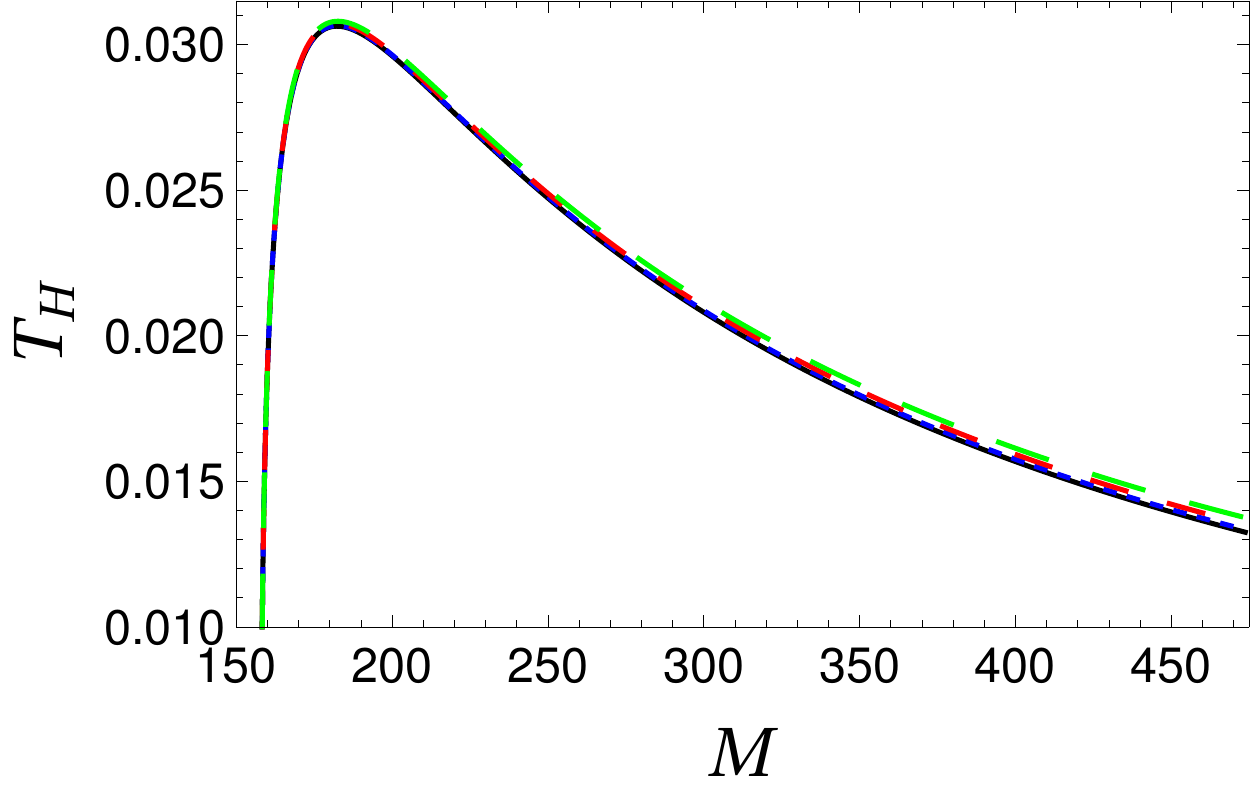}   \
\includegraphics[width=0.32\textwidth]{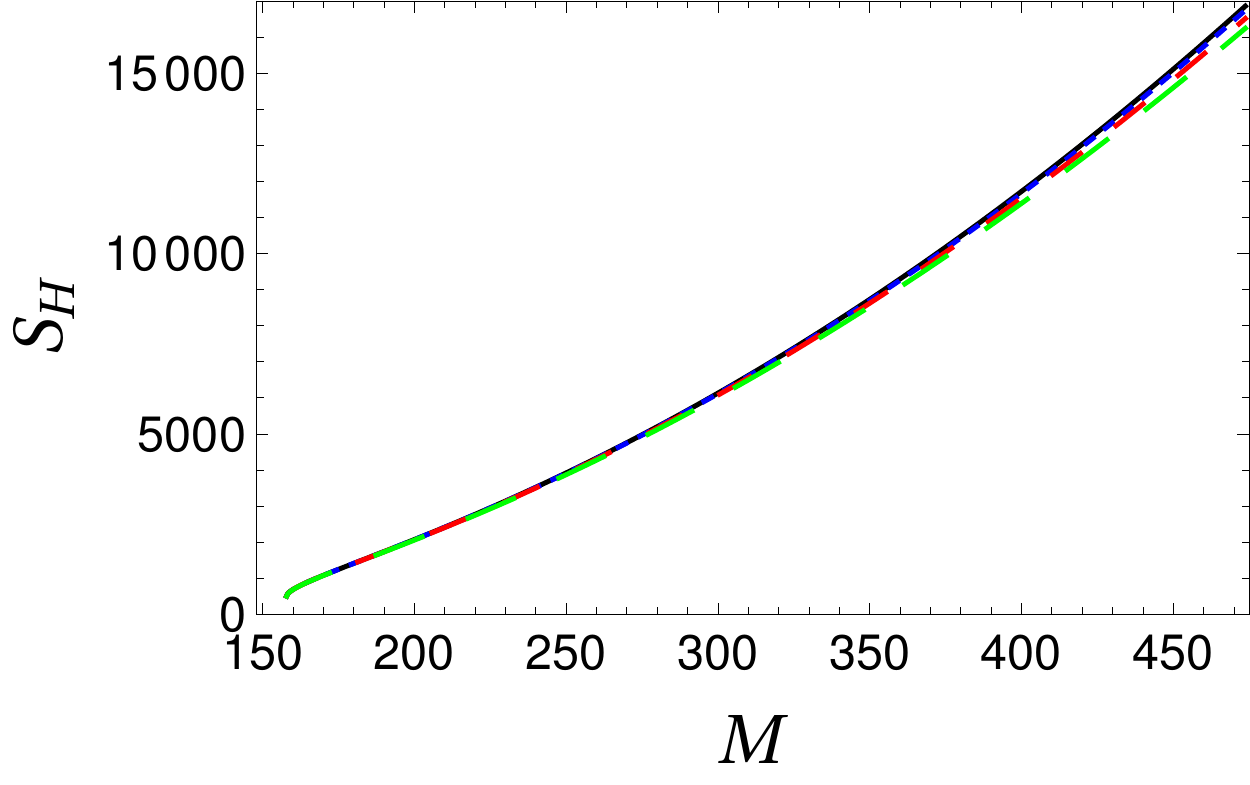}   
\\
\includegraphics[width=0.32\textwidth]{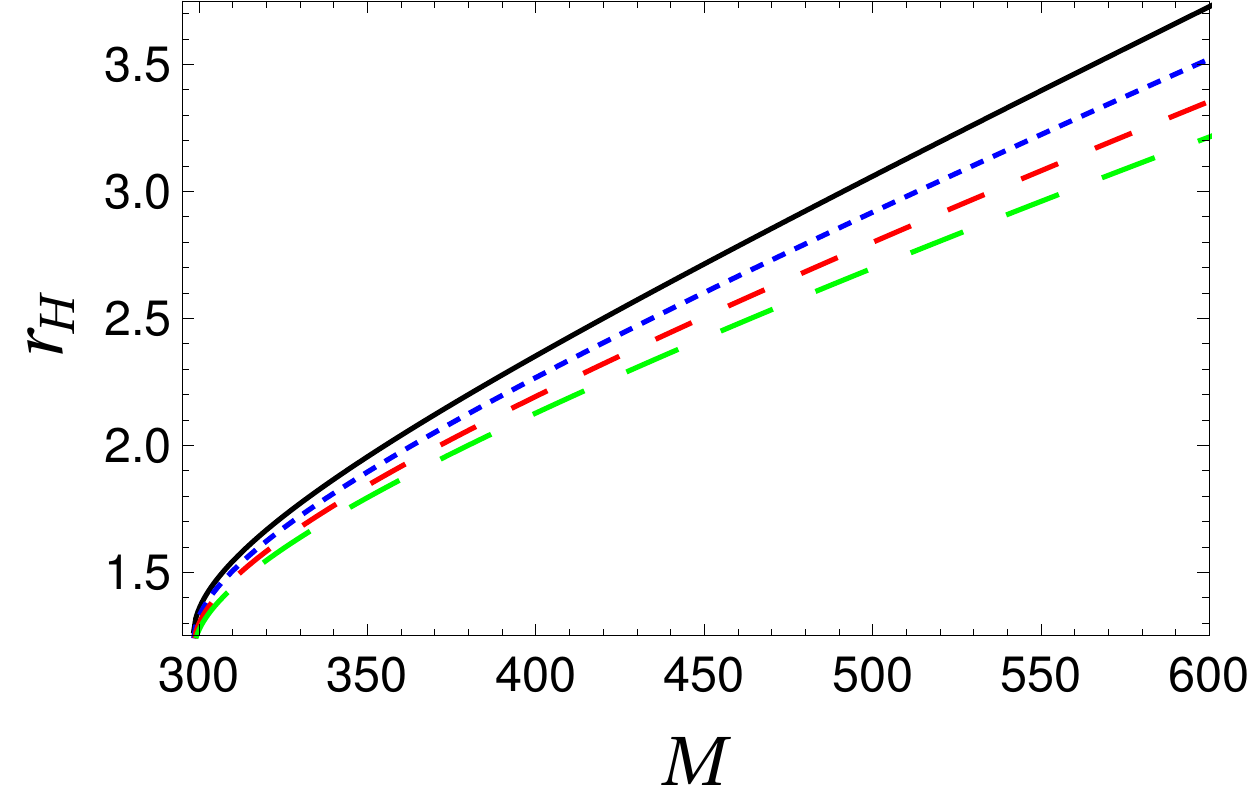}   \
\includegraphics[width=0.32\textwidth]{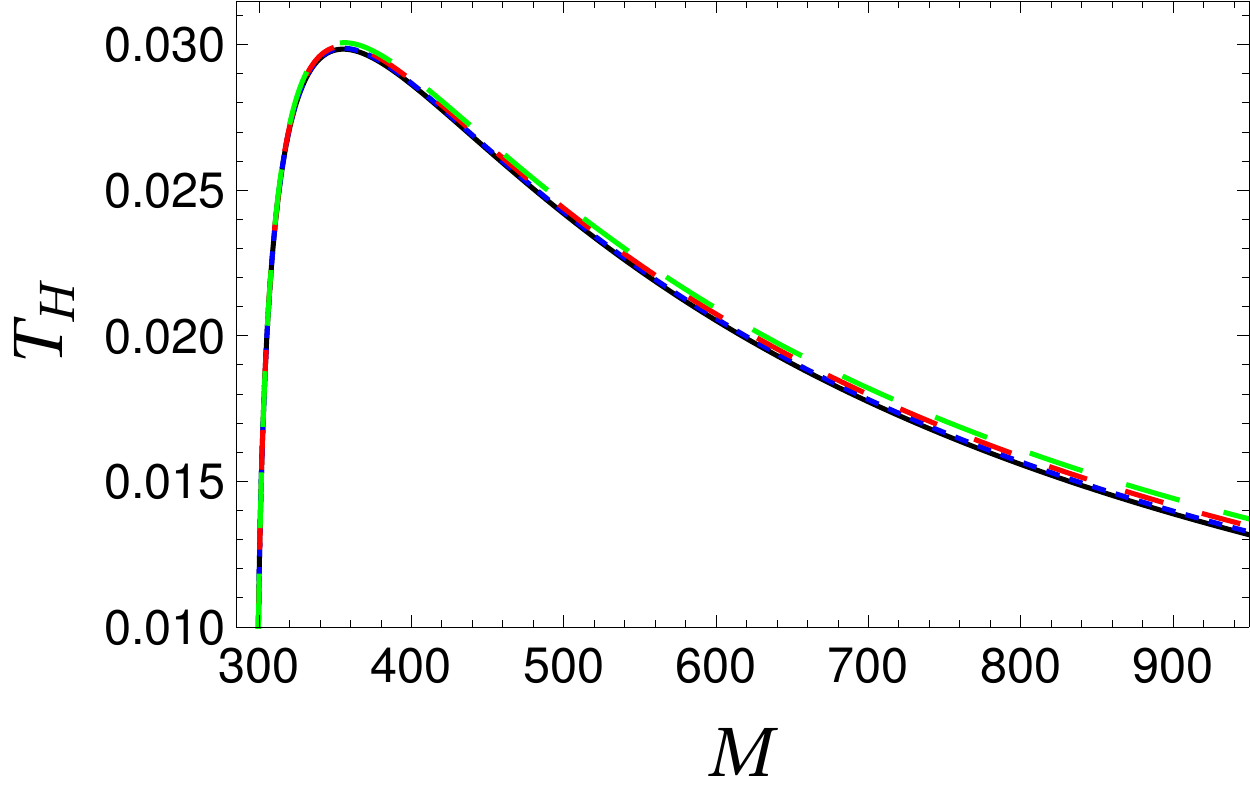}   \
\includegraphics[width=0.32\textwidth]{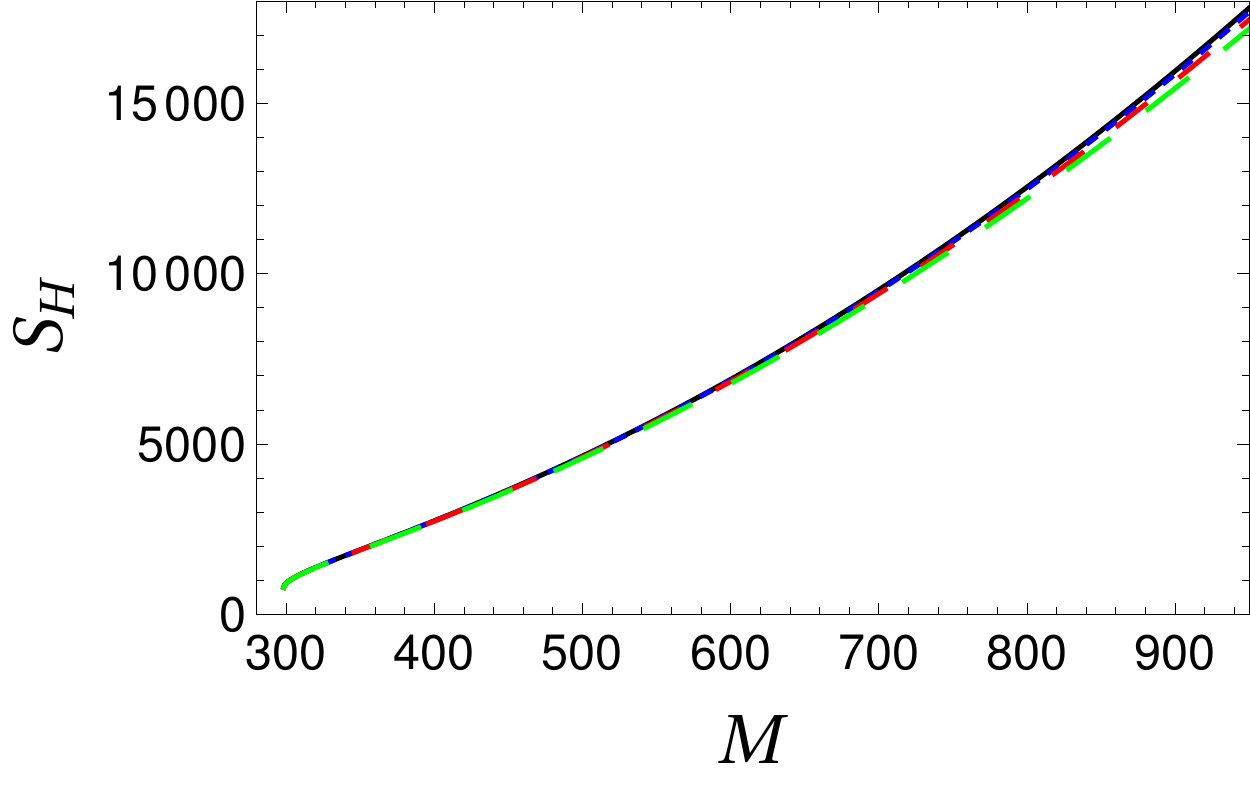} 
\caption{
{\bf{Left:}} Scale-dependent black hole horizon $r_H$ versus the classical black hole mass, 
{\bf{Middle:}} Hawking temperature against the black hole mass, and, finally,
{\bf{Right:}} Bekenstein-Hawking entropy versus the classical black hole mass.
The first column correspond to the horizon, the second line correspond to Hawking temperature and finally, $S_H$ for the right hand side. 
The first (left),  second (middle) and third (right) column correspond to the cases $\alpha= \{2,3\}$ respectively. We show the classical model (solid black line) and three different cases for each figure: 
i) $\epsilon = 0.033$ (dashed blue line), ii) $\epsilon = 0.067$  (dotted red line) and iii) for $\epsilon = 0.010$  (dotted dashed green line). 
We have used the set 
$\{Q_0, e_0, G_0, M_0\} = \{ 1, 1/(2\sqrt{\pi}), 1/(16 \pi^2), 32 \pi^2 \}$ .
%
}
\label{fig:2}
\end{figure*}

The Bekenstein-Hawking entropy in SD gravity can be safely computed by considering the theory as a special subclass of scalar-tensor theories. 
As it is well known from Brans-Dicke theory \cite{Jacobson:1993vj,Iyer:1995kg}, the entropy of black hole solutions in $d + 1$ spacetime dimensions with varying Newton's constant is computed as follow
\begin{align}
S &= \frac{1}{4}\oint_{r=r_H} \mathrm{d} ^{d-1}x \frac{\sqrt{h}}{G(x)},
\end{align}
where $h_{ij}$ is the induced metric at the horizon $r_H$. For the present spherically symmetric solution this integral is
quite simple. 
So, $G(x) = G(r_H)$ is constant along the horizon due to spherical symmetry.
Thus, for practical purposes, it is enough to replace $G_0 \rightarrow G(r)$ to obtain
\begin{align}
S_H &= \frac{\mathcal{A}_H}{4 G_0}(1 + \epsilon r_H).
\end{align}
Thus, for large values of the combination $\epsilon r_H$, the corresponding entropy is considerably disturbed; otherwise, the corrections are practically imperceptible. In this case, we still require a concrete form of the BH horizon. Given that we cannot analytically find it, we again take advantage of the numerical solutions. Fig.~\eqref{fig:2} (right panel) shows, from top to down, the Bekenstein-Hawking entropy for $\alpha = 2$ and $\alpha=3$ (down) for several values of the running parameter, $\epsilon$. Notice that the entropy is smaller than its classical counterpart, which we think is remarkable. 
The heat capacity can be obtained from the following definition:
\begin{align}
C_H \equiv T \ \frac{\partial S}{\partial T} \Biggl |_Q ,
\end{align}
and simplifying, we finally write the compact expression
\begin{align}
C_H = - S_0(r_H)(1 + \epsilon r_H).
\end{align}
Interestingly, the negative sign is maintained in the SD version of the EpM BH solution. Thus, in this sense, the BH is still unstable when Newton's coupling is positive. As the combination $\epsilon r_H$ is always small, the potential corrections to the heat capacity are quite weak.

\section{Conclusions}

To summarize, in the present work, we have discussed a charged black hole solution in four-dimensional spacetime in light of the scale-dependent scenario, strongly inspired by asymptotically safe gravity. After a short pass by the effective action, we have derived the corresponding effective Einstein's field 
equations coupled to non-linear electrodynamics.
 Then, we have computed the metric potential as well as the basic thermodynamics properties. 
 We have observed that all new non-classical effects are controlled by the running parameter, $\epsilon$. 
 Thus, our solution mimics the classical one when $\epsilon \sim 0$ and differences emerge when $\epsilon$ becomes large. Finally, we have pointed out that, to obtain a well-defined solution, the black hole mass should satisfy a minimum value which is present both in the classical and in the scale-dependent settings.

\section*{Acknowledgements}

We are grateful to the anonymous reviewer for a careful reading of the manuscript as well as for numerous useful comments and suggestions. The authors wish to thank C.~Herdeiro for correspondence.
The author A.~R. acknowledges DI-VRIEA for financial support through Proyecto Postdoctorado 2019 VRIEA-PUCV.
P. B. is funded by the Beatriz Galindo contract BEAGAL 18/00207 (Spain).
The author G.~P. thanks the Funda\c c\~ao para a Ci\^encia e Tecnologia (FCT), Portugal, for the financial support to the Center for Astrophysics and Gravitation-CENTRA, Instituto Superior T\'ecnico, Universidade de Lisboa, through the Grant No. UID/FIS/00099/2013.



\bibliographystyle{unsrt}         

\end{document}